\documentclass[prc,showpacs,twocolumn,superscriptaddress]{revtex4}
\usepackage{graphicx}
\usepackage{dcolumn}
\usepackage{bm}

\begin{document}

\title{Least action description of spontaneous fission in fermium and 
nobelium nuclei based on the Gogny energy density functional}

\author{R.~Rodr\'{\i}guez-Guzm\'an}

\affiliation{Physics Department, Kuwait University, 13060 Kuwait, Kuwait}

\author{L.~M.~Robledo}

\affiliation{Center for Computational Simulation,
Universidad Polit\'ecnica de Madrid,
Campus de Montegancedo, Boadilla del Monte, 28660-Madrid. Spain}

\affiliation{Departamento de F\'\i sica Te\'orica, Universidad
Aut\'onoma de Madrid, E-28049 Madrid, Spain}

\date{\today}

\begin{abstract}
The systematic of the spontaneous fission half-lives for the nuclei 
$^{242-262}$Fm and $^{250-260}$No is analyzed, within a least action scheme,
with the parametrization D1M of the Gogny energy density functional.
The properties of the dynamic (least action) fission 
paths are analyzed and compared to those of the static (minimal energy) 
ones. The constrained Hartree-Fock-Bogoliubov approximation is used to 
compute deformed mean-field configurations, zero-point quantum 
corrections and collective inertias. It is shown that a cumbersome full 
variational search of the least action fission path, within the space 
of HFB states, might not be required if the relevant degrees of freedom 
are taken into account in the minimization of the 
Wentzel-Kramers-Brillouin action. The action is minimized in terms of 
pairing fluctuations that explore the pairing content of the HFB states 
along the fission paths of the considered nuclei. It is found that, for 
a given shape, the minimum of the action in fermium and nobelium nuclei 
corresponds to a value of the pairing fluctuations larger than the one 
associated with the minimal energy solution for the same shape. The 
reduction of the action, via larger pairing correlations, has a 
significant impact on the predicted spontaneous fission half-lives 
improving their comparison with the experiment by several orders of 
magnitude.  
\end{abstract}


\maketitle{}

%
%
%

\section{Introduction}

Disentangling which are the most relevant degrees of freedom in the 
fission of the atomic nucleus still remains  a major challenge in 
today's nuclear structure physics 
\cite{Specht,Bjor,Krappe,Schunck2016}. Fission is traditionally 
portrayed as the smooth evolution of the nuclear shape from the one of 
the ground state to the shape at scission. Shape evolution is governed 
by both subtle quantum mechanics effects and the competition of global 
properties of the nuclear interaction like surface properties and 
Coulomb repulsion. The dynamics of shape evolution is usually described 
in terms of a  Hartree-Fock-Bogoliubov (HFB) mean field with
constrains on some deformation parameters {\bf{Q}} like the quadrupole 
moment, hexadecapole moment or necking. The evolution of the energy 
with deformation leads to a potential energy curve (or surface, in the 
general case) (PES) that, along with some collective inertias, can be 
used to determine fission observables like spontaneous fission 
lifetimes ($t_{SF}$) which are traditionally computed within the 
Wentzel-Kramers-Brillouin (WKB) framework used to describe tunneling through a 
barrier. Super-heavy nuclei ($Z>100$) belong to the class of nuclei 
where there is no classical fission barrier and therefore only quantal 
effects are responsible for their stability. Therefore, they are the 
perfect laboratory to understand the subtle quantal effects relevant in 
the dynamics of fission \cite{Baran2015}. The insight gained on the 
properties of those quantal effects can provide key information on the 
very limits of nuclear stability and the existence of super-heavy 
nuclei beyond Oganesson \cite{naza2018}. Among super-heavy nuclei, extensive 
experimental studies \cite{Holden-tsf-exp} have been able to determine the spontaneous 
fission lifetime of a wealth of isotopes of fermium and nobelium. 
On the other hand, the reproduction of the 
inverted parabola behavior of $t_{SF}$ as a function of neutron number 
characteristic of both isotopic chains has been a challenge for 
microscopic mean field models. For these reasons, we have chosen the two 
mentioned isotopic chains as the subject of our present study.

In recent years, the constrained mean-field approximation \cite{rs} has 
emerged as a useful tool to study the gross features of fission from a 
microscopic perspective \cite{Schunck2016}. Calculations are typically 
carried out with (effective) non-relativistic interactions like Gogny 
\cite{gogny-d1s,Delaroche-2006,Robledo-Martin,Dubray,PEREZ-ROBLEDO,Younes2009,WERP02,Egido-other1, 
Warda-Egido-2012}, Skyrme 
\cite{UNEDF1,Mcdonell-2,Erler2012,Baran-SF-2012,Baran-1981}, and 
Barcelona-Catania-Paris-Madrid (BCPM) 
\cite{BCPM,Robledo-Giulliani,Giuliani2018} or with relativistic 
\cite{Bender-1998,Abusara-2010,Abu-2012-bheights,RMF-LU-2012,Kara-RMF} 
energy density functionals (EDFs).  One of the main outcomes of those 
mean-field calculations is the potential energy surface given as a 
function of  quadrupole, octupole, $\dots$, or necking deformation 
parameters {\bf{Q}} = ($Q_{20}$, $Q_{22}$, $Q_{30}$, $\dots$, 
$Q_{Neck}$) and determined by self-consistent HFB constrained calculations. 
Moreover, the mean-field framework also provides the 
associated collective inertias as well as the quantum zero-point  
rotational and vibrational energy corrections. All those ingredients 
are required to compute fission observables like, for example, the 
spontaneous fission  half-lives $t_{SF}$. Some studies 
\cite{Rutz1997,Bender-1998,Buervenich2004,Baran2015} comparing the 
predictions of the different interactions just mentioned come to the 
conclusion that the differences found  among them are at the 
quantitative, not the qualitative, level.


\begin{figure}
\includegraphics[width=0.46\textwidth]{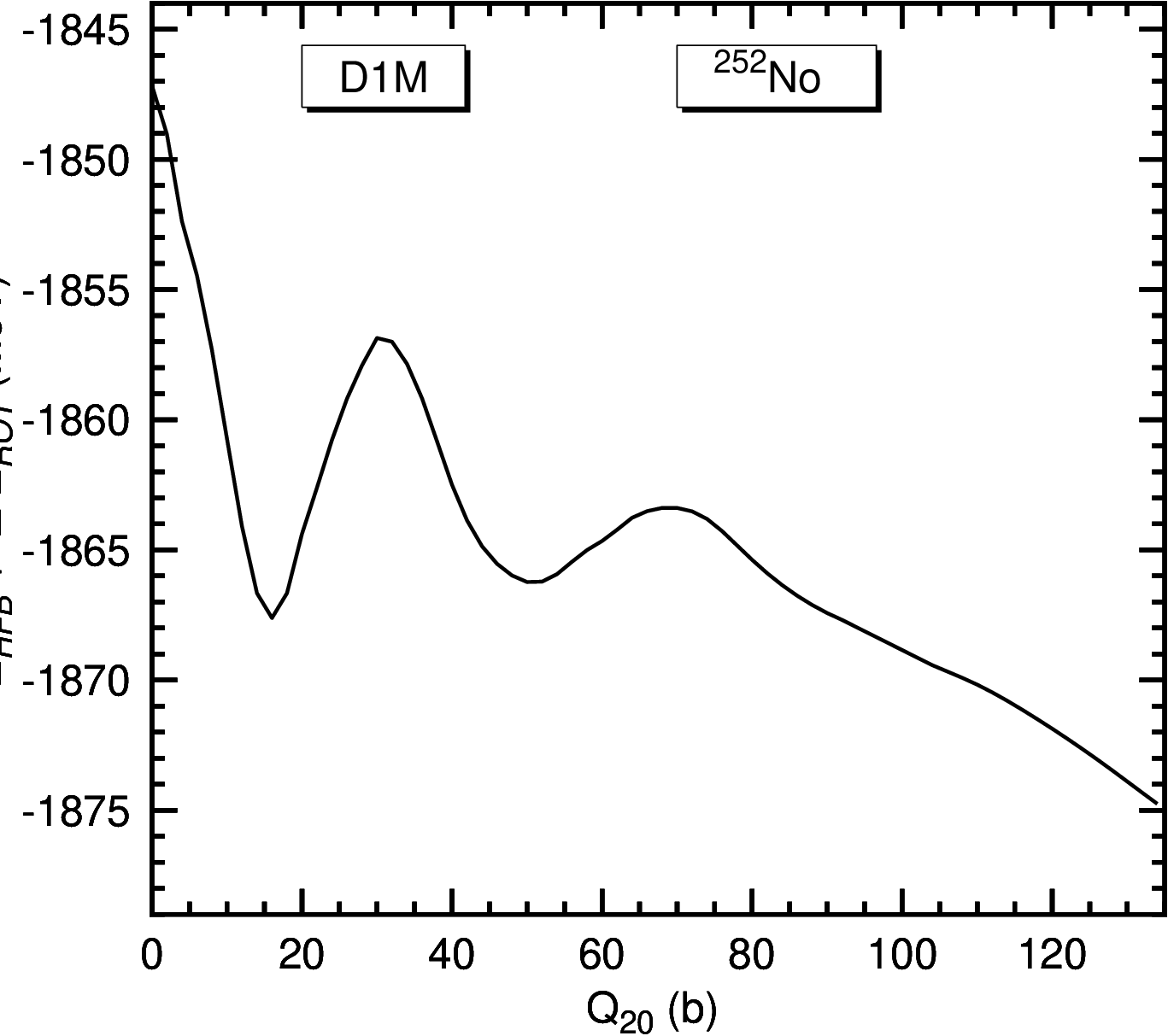}
\caption{The HFB plus rotational correction energy
corresponding to the static fission path  is
plotted, as a function of the quadrupole 
moment $Q_{20}$, for the nucleus $^{252}$No.}\label{peda-252No} 
\end{figure} 


Within the standard  approach, the HFB wave functions along the 
fission path are obtained by minimizing the constrained HFB energy in 
accordance to the  Ritz-variational principle \cite{rs}. In the last 
few years we have resorted to this approach to describe the fission 
properties of even-even Ra, U and Pu nuclei as well as for a selected 
set of super-heavy elements \cite{Rayner-U,Rayner-Pu,Rayner-Ra} using 
the three most important parametrizations of the Gogny-EDF, namely, D1M \cite{gogny-d1m}, D1N 
\cite{gogny-d1n} and D1S \cite{gogny-d1s}. 
Special attention has also been paid to understand the uncertainties in the 
predicted  $t_{SF}$ values arising from the different building blocks 
entering the semi-classical WKB formula used to compute them.  For 
instance, it has  been found in those studies that modifications of 
a few per cent in the pairing strengths can have a significant impact 
on the collective masses leading to uncertainties of several orders of 
magnitude in the predicted $t_{SF}$ values. Though there is a large 
variability of the predicted $t_{SF}$ values with respect to the 
details involved in their computation, it has also been shown that this 
mean field framework  produces a robust trend with neutron number for 
the considered even-even nuclei and the changes in the different 
building blocks only amount to a parallel displacement of the curves. 
Similar conclusions can be extracted from recent fission studies for 
odd-mass U, Pu and No nuclei \cite{Rayner-No,Rayner-U_Pu-odd} within 
the Equal Filling Approximation (EFA) \cite{Sara-Robledo_EFA}.

The calculations mentioned above belong to the so called "static" category where
the guiding principle is the least energy principle. This is 
considered to be just an approximation to the "least action" principle,
known as the "dynamic" approach, involving the minimization of
the action associated with a given set of relevant coordinates.
It is thought that the dynamic approach 
is better suited to describe the tunneling through the barrier. 
It is well known that  both the  static and dynamic approaches provide similar results  
\cite{Min_Action_Skyrme_1,Min_Action_Gogny} when considering only shape degrees 
of freedom. This is the traditional 
justification for using the simpler static approach. However, the key 
role played by pairing degrees of freedom in the dynamics of fission was already 
pointed out in Ref.~\cite{Moreto-pairing-fission}. It was found 
that the competition between the collective inertia (decreasing as the 
inverse of the square of the pairing gap 
\cite{proportional-1,proportional-2}) and  the energy (increasing as 
the square of the pairing gap)  leads to a minimum of the action at a 
larger pairing gap than the one corresponding to the minimum energy 
approach. A number of spontaneous fission studies have already 
considered the effect of pairing fluctuations in fission 
\cite{Urin,Lublin-1,Lublin-2,Sta-P2,Sad14,Min_Action_Skyrme_2,Min_Action_RMF} 
showing that including pairing as a dynamical variable produces 
a reduction of a few orders of magnitude in the computed $t_{SF}$ 
values as compared to the ones obtained within the static approach. 
Moreover, pairing fluctuations can also restore axial symmetry along 
the fission path \cite{Sad14,Min_Action_Skyrme_2}. 


\begin{figure} 
\includegraphics[width=0.46\textwidth]{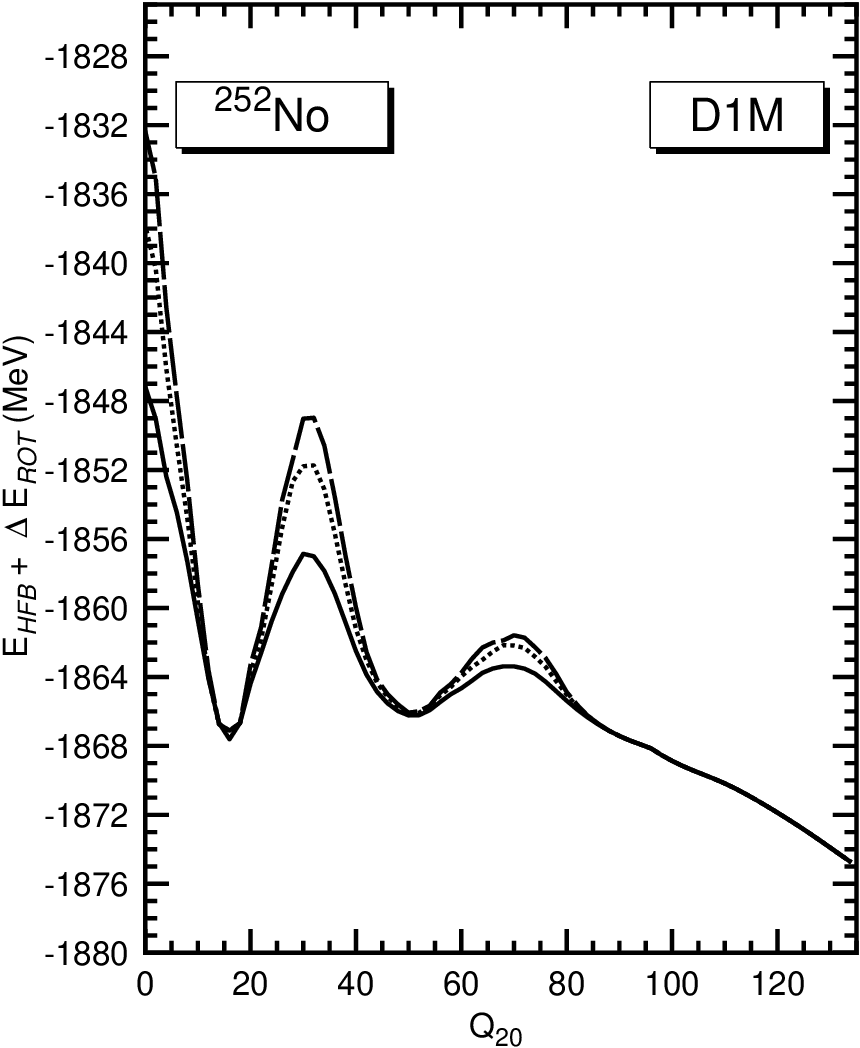} 
\caption{The HFB plus rotational correction energy
corresponding to the dynamic ATDHFB 
(dashed line)
and GCM (dotted line) fission paths are 
plotted, as functions of the quadrupole moment $Q_{20}$, for the nucleus 
$^{252}$No. The energies corresponding to the static path (full line)  are also 
included in the plot. For details, see the main text. } 
\label{paths-252No} 
\end{figure} 

Another observable of interest is the fission fragments' mass distribution.
This quantity can be studied microscopically within time dependent
approximations like the Time Dependent Generator Coordinate Method
(TDGCM) \cite{Ber91,Gou05,Reg16,Tao17,Zdeb17}, stochastic time dependent
HF+BCS \cite{Tan17}  or the Time Dependent HFB (TDHFB) \cite{God15,Bul16,Sim18} 
requiring enormous computational
resources. It has been shown \cite{Tao17} that increasing the pairing
strength by a few percent modifies in a substantial way the fragments' mass
distribution, showing the relevance of the pairing degree of freedom also
in this respect. Therefore, it would be very interesting to extend the 
present framework, based on the least action principle, to the time 
dependent case, but this is out of the scope of this paper. 


Recently, we have studied the predictions for spontaneous fission 
half-lives within a least action scheme based on the Gogny-D1M EDF for 
a selected set of U isotopes \cite{Min_Action_Gogny}. On the one hand, 
our calculations corroborate that the static and dynamic approaches 
lead to similar results when shape degrees of freedom are the only ones 
considered. On the other hand, the results show the large impact on the 
action coming from the particle number fluctuation $\langle \Delta 
\hat{N}^2 \rangle$ degree of freedom (a quantity connected with pairing correlations). 
For a given nuclear shape, labeled by the quadrupole moment $Q_{20}$, 
the minimum of the action, in the direction of the  $\langle \Delta 
\hat{N}^2 \rangle$ variable, can be up to a factor of three smaller 
than the value of the action for the minimum energy state with the same 
$Q_{20}$ value. This reduction has a strong impact on the predicted 
$t_{SF}$ values and dramatically improves the agreement with the 
experiment in the case of  $^{232-238}$U.


\begin{figure*} 
\includegraphics[width=1.0\textwidth]{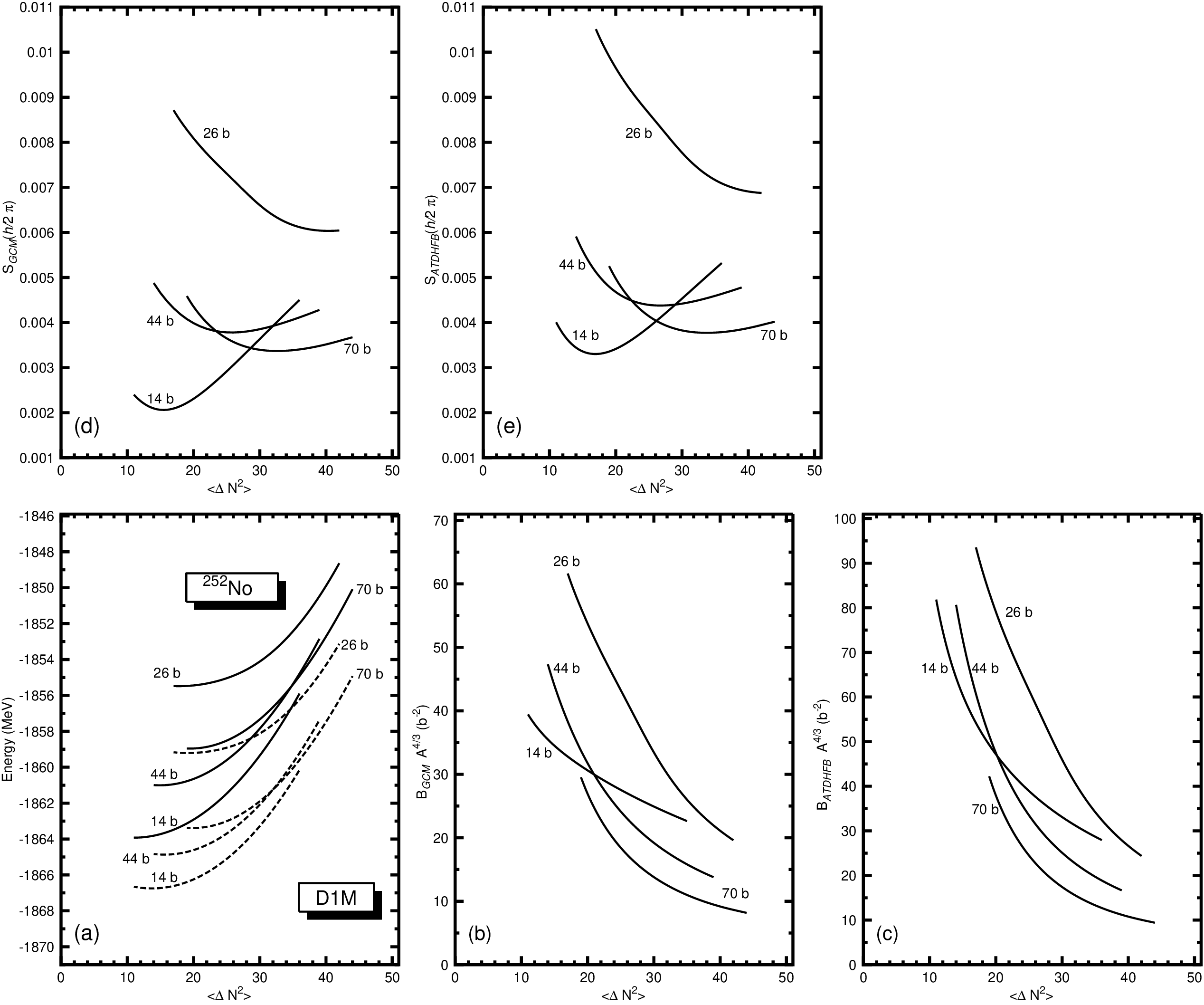} 
\caption{The intrinsic HFB energies (full lines) and 
the HFB plus  rotational correction energies (dashed lines) [panel 
(a)], GCM and ATDHFB collective masses [panels (b) and (c)] as well as 
the GCM and ATDHFB integrand of the action Eq.(\ref{Action}) [panels 
(d) and (e)], are plotted  as functions of the particle number 
fluctuations $\langle \Delta \hat{N}^2 \rangle$. Results are shown for 
the quadrupole moments $Q_{20}$ = 14, 26, 44 and 70 b. For more 
details, see the main text. } \label{fig-conjunto-E-BATD-BGCM} 
\end{figure*}

In this paper we 
consider again the  particle number fluctuation $\langle \Delta \hat{N}^2 
\rangle$ as the relevant degree of freedom in the minimization of the 
action and use this framework to compute the spontaneous fission half-lives for 
the super-heavy isotopes $^{242-262}$Fm  and $^{250-260}$No. The choice 
is driven by the existence of experimental data for those isotopes 
\cite{Holden-tsf-exp} and also because they belong to a 
region of the nuclear chart where the shell effects 
characteristic of super-heavy elements with $Z>100$ start to manifest  
\cite{Warda-Egido-2012,Rayner-No}. Moreover,  accounting for 
the experimental bell-shaped dependence of the spontaneous fission 
half-lives as functions of the neutron number \cite{Holden-tsf-exp} in 
fermium and nobelium nuclei represents a very challenging test for any model
of fission.

In the calculations we have used the Gogny-D1M  EDF \cite{gogny-d1m}. 
The suitability of the Gogny-D1M EDF to describe fission in heavy 
and super-heavy nuclei has been demonstrated in previous studies 
\cite{Rayner-U,Rayner-Pu,Rayner-Ra,Rayner-No,Rayner-U_Pu-odd} 
where the results for barrier heights, 
excitation energies of fission isomers and half-lives 
have shown a good agreement with experimental data. The comparison
with other theoretical studies using other parametrizations 
\cite{Delaroche-2006,Robledo-Giulliani,WERP02,Warda-Egido-2012} 
is also satisfactory. This is a nice feature as the parametrization 
D1M does a much better job in reproducing nuclear masses \cite{gogny-d1m} 
than D1S. In addition, D1M performs as well as D1S in the description of 
nuclear structure phenomena 
\cite{PRCQ2Q3-2012,PTpaper-Rayner,Rayner-Sara,Rayner-Robledo-JPG-2009,Robledo-Rayner-JPG-2012,Rayner-PRC-2010,Rayner-PRC-2011}.

The paper is organized as follows. In Sec.~\ref{Theory}, we briefly 
outline the procedure to find the least action path using the 
quadrupole moment $Q_{20}$ and particle number fluctuation $\langle 
\Delta \hat{N}^2 \rangle$ as relevant coordinates. The results of our 
calculations are discussed in Sec.~\ref{results}. First, in Sec. 
\ref{strategy-252No}, we illustrate the method in the case of 
$^{252}$No. The systematic of the  spontaneous fission half-lives in 
$^{242-262}$Fm  and $^{250-260}$No and the properties of the least 
action  paths are presented in Sec.~\ref{systematics-paths} where we 
also compare with results obtained within the static framework. 
Finally, Sec.~\ref{Coclusions} is devoted to the concluding remarks and 
work perspectives. 


\section{Theoretical framework} 
\label{Theory} 
 
The mean-field approximation  based on (product) HFB wave functions \cite{rs} 
has been used in the present study. Constrains in 
the mean value of the axially symmetric quadrupole $\hat{Q}_{20}$ and  
octupole $\hat{Q}_{30}$ operators as well as on the particle number 
fluctuations $\Delta \hat{N}^2$  \cite{Min_Action_Gogny} have been 
used. Though it will not be mentioned explicitly, in all the HFB 
calculations discussed below, aside from the usual  constrains on both 
the proton and neutron numbers \cite{rs}, a constrain on the operator 
$\hat{Q}_{10}$ is used  to prevent spurious effects associated to the 
center of mass motion. Note, that parity is allowed to be broken at any 
stage of the calculations as required by the physics of asymmetric 
fission.

The HFB quasiparticle operators,  have been expanded in an axially 
symmetric (deformed) harmonic oscillator  (HO) basis containing  states 
with $J_{z}$ quantum numbers up to 35/2 and up to 26 quanta in the z 
direction. They correspond to those satisfying the condition $ 2 
n_{\perp} + |m| + q\, n_{z} \le M_{z, \mathrm{MAX}} $ with $M_{z, 
\mathrm{MAX}}=17$ and $q=1.5$. This choice is well suited for the 
elongated prolate shapes  typical of the fission process 
\cite{Robledo-Giulliani,WERP02}. The two HO 
lengths  $b_{z}$ and $b_{\perp}$ have been optimized so as to minimize 
the total HFB energy for each value of the quadrupole moment. The 
computationally expensive oscillator length optimization along with the 
large HO basis used guarantees good convergence in the potential energy 
surface in the relevant range of shape deformation 
\cite{WERP02}. An approximate second order 
gradient method \cite{Robledo-Bertsch2OGM} has been used for the 
solution of the HFB equation. The main reason for this choice is the 
clear advantage  of this type of methods over the more traditional 
successive diagonalization method  when many constrains are imposed. 
As it is customary in all the Gogny force parametrizations, the 
two-body kinetic energy correction, including the exchange and pairing 
channels, has been taken into account in the Ritz-variational  
procedure. On the other hand, the Coulomb exchange term is considered 
in the Slater approximation \cite{CoulombSlater} while the Coulomb and 
spin-orbit contributions to the pairing field have been neglected.

Within the WKB formalism  the $t_{SF}$ half-life (in seconds) is given by
\begin{eqnarray} \label{TSF-WKB}
t_{SF} = 2.86 \times 10^{-21} \times \left(1 + e^{2S} \right)
\end{eqnarray}
where the action $S$ along the (one-dimensional $Q_{20}$-projected) 
fission path reads
\begin{eqnarray} \label{Action}
S = \int_{a}^{b} dQ_{20} {\cal{S}}(Q_{20})
\end{eqnarray}
and  the integrand ${\cal{S}}(Q_{20})$ takes the form
\begin{eqnarray} \label{Integrand-Action}
{\cal{S}}(Q_{20}) = \sqrt{2B(Q_{20})\left(V(Q_{20})-\left(E_{Min}+E_{0} \right)  \right)}
\end{eqnarray}

The integration limits $a$ and $b$ are the classical turning points 
\cite{proportional-1} for the potential $V(Q_{20})$  corresponding to 
the energy $E_{Min}+E_{0}$. The energy $E_{Min}$ corresponds to the 
ground state minimum for the considered path while $E_{0}$ accounts for 
the true ground state energy once quadrupole fluctuations are taken 
into account. In this work, we have taken the typical  value $E_{0}$ = 
0.5 MeV \cite{WERP02,Warda-Egido-2012}. In 
Eq.(\ref{Integrand-Action}), $B(Q_{20})$ represents the collective 
mass. The potential  $V(Q_{20})$ is given by the HFB energy corrected 
by the zero-point vibrational and rotational energies. For the 
evaluation of the collective mass  and the zero-point vibrational 
energy correction  two methods have been used. One is the cranking 
approximation \cite{crankingAPPROX,Giannoni,Libert-1999} to the 
Adiabatic Time Dependent HFB (ATDHFB) scheme. The second method is 
based on the Gaussian Overlap Approximation (GOA)  to the GCM. 
Details on how to compute the required  quantities can be 
found, for instance, in Refs.~\cite{Krappe,Schunck2016}.  As it is customary 
in fission calculations, the coupling with other degrees of freedom in 
the evaluation of the collective mass is neglected within both the 
ATDHFB and GCM schemes \cite{Min_Action_Gogny} and only the quadrupole 
inertia is considered. The  rotational energy correction 
$\Delta E_\mathrm{ROT}$ has been computed in terms of the Yoccoz 
moment of inertia \cite{RRG23S,ER-Lectures,NPA-2002}.


\begin{figure} 
\includegraphics[width=0.46\textwidth]{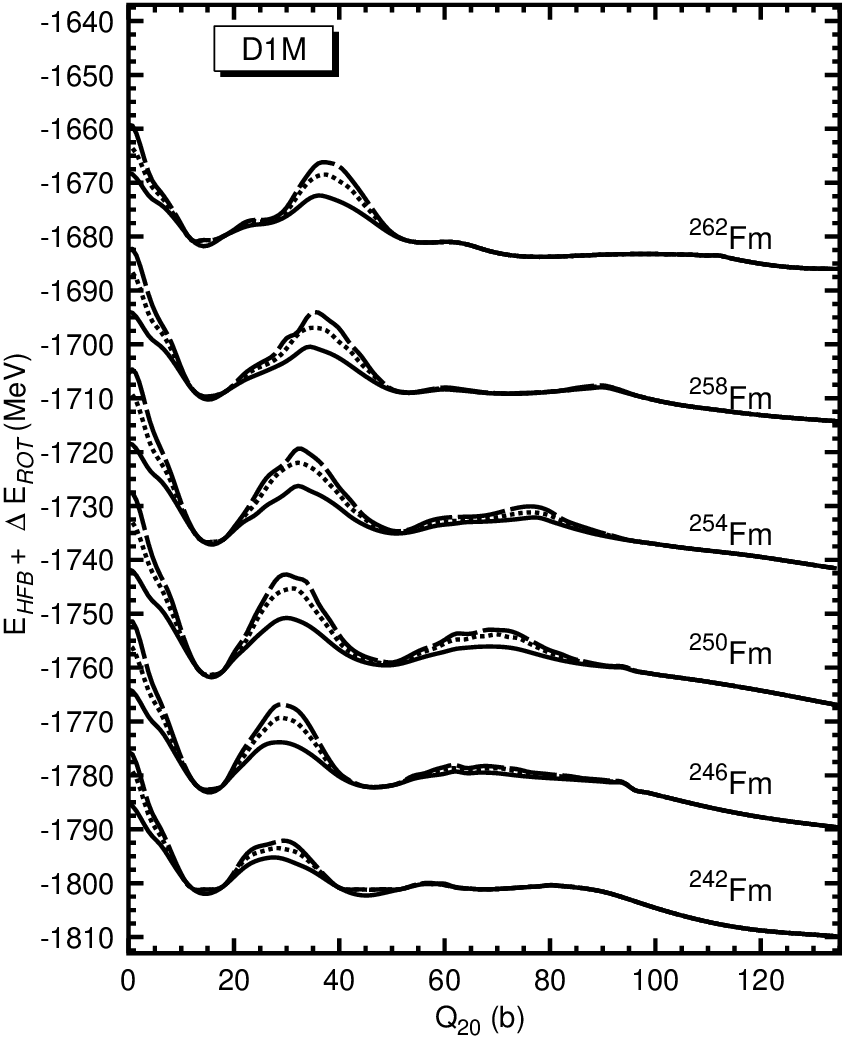} 
\caption{The HFB plus  rotational correction energies 
corresponding to the dynamic ATDHFB (dashed line) and GCM 
(dotted line)
fission paths are plotted  
as functions of the quadrupole moment $Q_{20}$ for the nuclei 
$^{242,246,250,254,258,262}$Fm. The energies corresponding to the 
static paths (full line) are also included in the plot. Starting from the nucleus 
$^{246}$Fm, the curves have been shifted by 50 MeV in order to 
accommodate them in a single plot. For details, see the main text. } 
\label{paths-242-246-250-254-258-260Fm} 
\end{figure}

In order to compute $t_{SF}$ using the least action framework we
have followed the procedure described below in order to find a good
approximation to the least action path. To illustrate the practical aspects of the 
methodology employed  we use the  nucleus $^{252}$No as an example.

\begin{description}

\item{Step 1:} 
We have first determined the least energy (i.e., static) fission path 
for $^{252}$No as a function of the axially symmetric quadrupole moment 
$Q_{20}$ using the HFB mean field procedure.  Zero-point quantum 
rotational and vibrational energies have been added {\it{a posteriori}} 
to the HFB energies. 

The static fission path obtained for $^{252}$No is depicted in  
Fig.~\ref{peda-252No} where the HFB plus  
rotational correction energy, is plotted as a function of the quadrupole moment.  
The ground state is located at $Q_{20}$ = 16 b and is reflection 
symmetric. The fission isomer at $Q_{20}$ = 52 b lies 1.38 MeV above 
the ground state from which, it is separated by the inner barrier 
($Q_{20}$ = 30 b) with the height of 10.75 MeV. As in the ground state 
case, the fission isomer is also reflection symmetric. 
Octupole correlations play a key role 
for quadrupole deformations $Q_{20} \ge$  62 b. Those correlations 
significantly affect the outer barrier ($Q_{20}$ = 70 b) whose height 
is 4.23 MeV. 

It should be kept in mind that the mean values of the hexadecapole 
$\hat{Q}_{40}$ and higher multipolarity operators are automatically 
given by the Ritz-variational procedure that determines the HFB  states 
$|\varphi(Q_{20}) \rangle$  corresponding to each of the intrinsic 
configurations along the static fission path shown in 
Fig.~\ref{peda-252No}. To each of the states  $|\varphi(Q_{20}) 
\rangle$ also corresponds a self-consistent value of the particle number 
fluctuations $\langle \Delta \hat{N}^2 \rangle_{self}$.

\item{Step 2:} 
For each of the different $Q_{20}$-configurations along the static path 
of $^{252}$No obtained in Step 1, we have performed a set of HFB 
calculations with constrains on both  $\Delta \hat{N}^2$ 
and $\hat{Q}_{20}$. The constrained value of $\Delta \hat{N}^2$ starts at the 
self-consistent value $\langle \Delta \hat{N}^2 \rangle_{self}$ and is 
increased until the minimum of the action of Eq. (\ref{Action}) is 
reached. We use a small step size and typically of the order of 20-30 
values of $\langle \Delta \hat{N}^2 \rangle$ are considered. In 
principle one should vary independently the particle number fluctuation 
for protons and neutrons but this would have a great impact on the 
computational cost of the calculation. We  have also checked that an 
independent variation of protons and neutrons do not bring much as 
compared to the variation of the total particle number fluctuation. 
This minimization search is performed for both the action obtained with 
the ATDHFB collective inertia as well as the one with the GCM 
collective mass. In this way we obtain two dynamically determined paths 
which have energies larger than the ones of the minimum energy path but 
for which the corresponding actions are minimized. By using this 
approximate way to find the minimum action configuration we avoid the use of sophisticated 
linear programming techniques required by the full minimization 
approach.
\end{description}


\begin{figure} 
\includegraphics[width=0.46\textwidth]{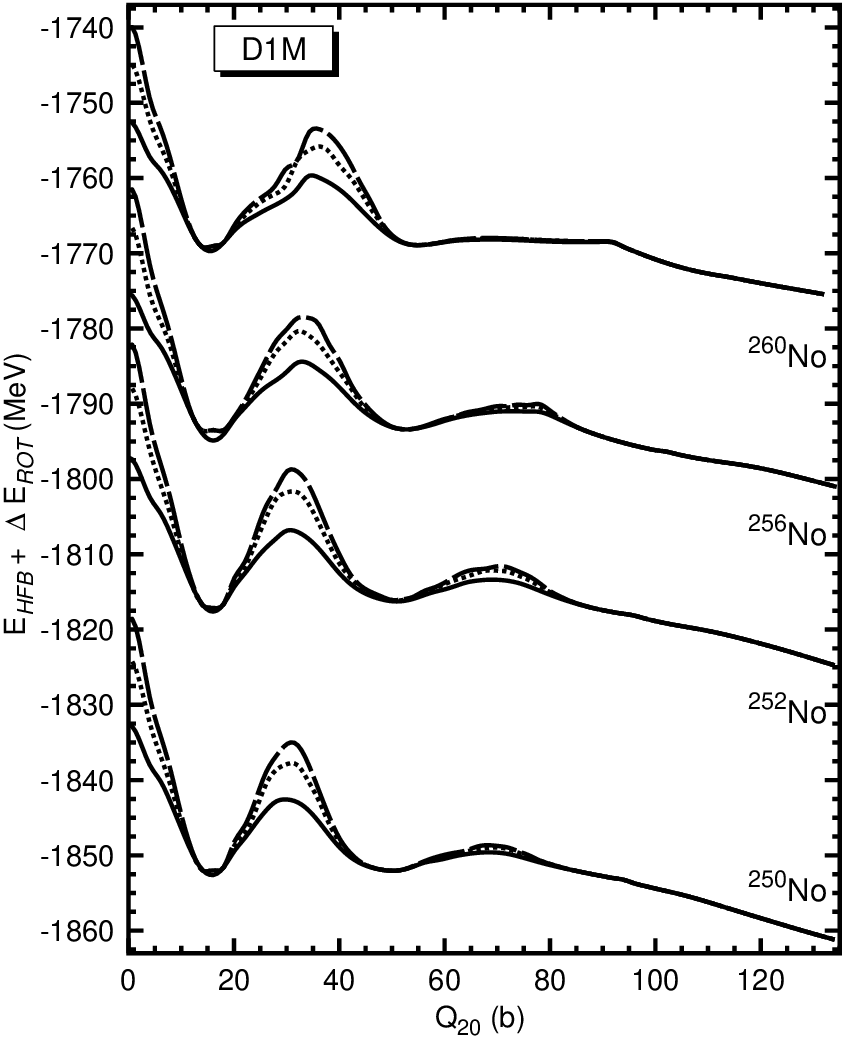} 
\caption{The HFB plus rotational correction energies 
corresponding to the dynamic ATDHFB (dashed line) and GCM 
(dotted line)
fission paths are plotted  
as functions of the quadrupole moment $Q_{20}$ for the nuclei 
$^{250,252,256,260}$No. The energies corresponding to the static paths
(full line) 
are also included in the plot. Starting from the nucleus $^{252}$No, 
the curves have been shifted by 50 MeV in order to accommodate them in 
a single plot. For details, see the main text. } 
\label{paths-248-252-256-260No} 
\end{figure}


\section{Discussion of the results}
\label{results}

In this section, we discuss the results of our calculations. First, in 
Sec.~\ref{strategy-252No}, we illustrate the methodology employed to 
compute the dynamic path in the case of  $^{252}$No. The same 
methodology is employed for all the studied nuclei. The systematic of 
the dynamic paths and spontaneous fission half-lives in $^{242-262}$Fm  
and $^{250-260}$No is presented in Sec.~\ref{systematics-paths}. 


\subsection{An illustrative example: the nucleus $^{252}$No}
\label{strategy-252No}

The HFB plus rotational correction energies corresponding to the 
dynamic ATDHFB and GCM fission paths are plotted  
as functions of the quadrupole moment $Q_{20}$ in Fig.~\ref{paths-252No} for the nucleus $^{252}$No. 
The energies corresponding to the static 
path are also included in the plot for the sake of comparison.

The deformations of the  absolute minimum ($Q_{20}$ = 16 b), the top of 
the inner barrier ($Q_{20}$ = 30 b), the fission isomer ($Q_{20}$ = 52 
b) and the top of the outer barrier ($Q_{20}$ = 70 b) corresponding to 
the dynamic path are similar to those for the static path. In the case 
of the  dynamic path, octupole correlations also play a key role for 
$Q_{20} \ge$  62 b. However,  the dynamic 
path displays larger inner and outer barrier heights  than the static 
one. The ATDHFB inner (outer) barrier height amounts to 18.09 (5.53) 
MeV while  the GCM inner (outer) barrier height turns out to be 15.38 
(4.94) MeV. These numbers should be compared with the values 10.75  and  
4.23 MeV for the inner and outer barrier heights of the static path. 
Large differences are also found at the spherical configuration lying 
34.78, 29.31 and 20.46 MeV above the absolute minima of the least 
action ATDHFB, GCM and the static paths, respectively. Obviously, both
kinds of barriers (static and dynamic) can not be compared as they come
from different approaches but the big differences observed raise the 
question about the suitability of comparing the minimum energy barrier heights
with the experimentally determined ones \cite{Bertsch2015}.


\begin{figure*} 
\includegraphics[width=1.\textwidth]{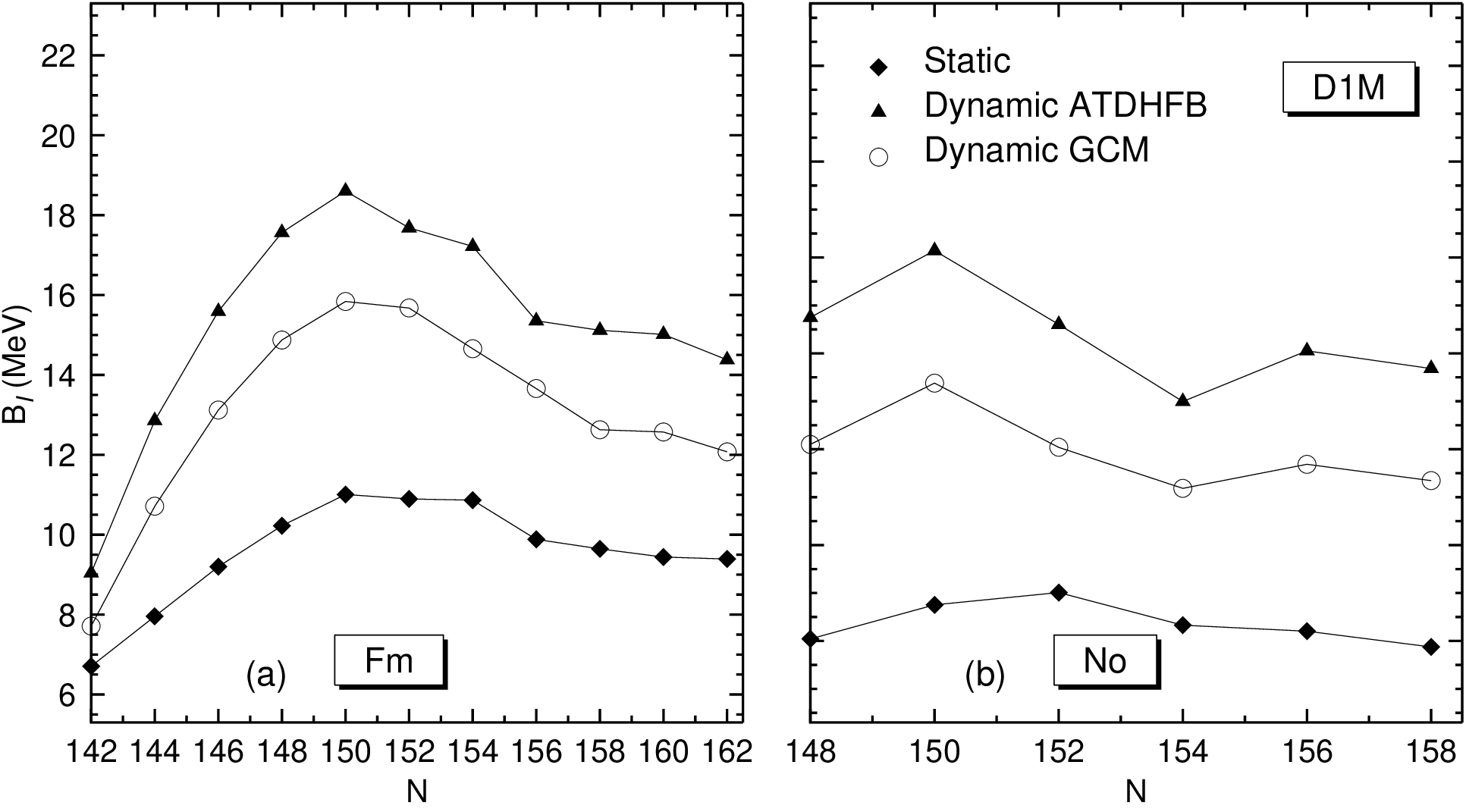} 
\caption{The  inner barrier heights corresponding to the 
dynamic GCM and ATDHFB fission paths in $^{242-262}$Fm  and 
$^{250-260}$No are plotted in panels (a) and (b), respectively,  as 
functions of the neutron number. The inner barrier heights 
corresponding to the static paths in those nuclei are also included in 
the plot. } \label{BI-syst-Fm-No} 
\end{figure*}

As  mentioned above (see, Sec.~\ref{Theory}), in order to minimize the 
action, we have carried out $\Delta \hat{N}^2$-constrained HFB 
calculations  for each $Q_{20}$-configuration along the static fission 
path of $^{252}$No. We have started at the self-consistent value 
$\langle \Delta \hat{N}^2 \rangle_{self}$  for each $Q_{20}$. As an 
example, in panel (a) of Fig.~\ref{fig-conjunto-E-BATD-BGCM}, we have 
plotted the intrinsic HFB energies (full lines) and the HFB plus the 
zero-point rotational energies (dashed lines) as functions of the 
particle number fluctuations $\langle \Delta \hat{N}^2 \rangle$ for 
relevant $Q_{20}$ values, namely 14, 26, 44 and 70 b. For example, for $Q_{20}$ = 44 b, 
$\langle \Delta \hat{N}^2 \rangle_{self}$ = 14 and we have considered 
14 $\le$ $\langle \Delta \hat{N}^2 \rangle$ $\le$ 39 with a mesh 
$\delta \langle \Delta \hat{N}^2 \rangle$ = 1. As can be seen, the 
intrinsic HFB energies exhibit an almost parabolic behavior as 
functions of $\langle \Delta \hat{N}^2 \rangle$ with a minimum at 
$\langle \Delta \hat{N}^2 \rangle$ = $\langle \Delta \hat{N}^2 
\rangle_{self}$. The same is also true for the HFB plus the 
zero-point rotational energy.

As it is well known, the ATDHFB masses are typically larger than the 
GCM ones 
\cite{Robledo-Giulliani,Baran-mass-2011,Rayner-U,Rayner-Pu,Rayner-Ra,Rayner-No,Rayner-U_Pu-odd}. 
As a consequence, the action in the exponent defining $t_{SF}$ 
Eq.(\ref{TSF-WKB}) is, in the ATDHFB case, larger than the GCM one. 
Depending on the value of the action, this difference can represent a 
change of several orders of magnitude in the predicted $t_{SF}$ 
values \cite{Rayner-U,Rayner-Pu,Rayner-Ra,Rayner-No,Rayner-U_Pu-odd}. 
This is the reason to consider both kinds of collective inertias in this 
study. In panels (b) and (c) of Fig.~\ref{fig-conjunto-E-BATD-BGCM}, we 
have plotted the GCM and ATDHFB inertias as functions of $\langle 
\Delta \hat{N}^2 \rangle$ for $Q_{20}$ = 14, 26, 44 and 70 b. 
Regardless of the numerical differences between both schemes, the 
collective masses decrease for increasing $\langle \Delta \hat{N}^2 
\rangle$. This agrees well with the inverse dependence of the 
collective masses on the square of the pairing gap 
\cite{proportional-1,proportional-2}.


\begin{figure} 
\includegraphics[width=0.46\textwidth]{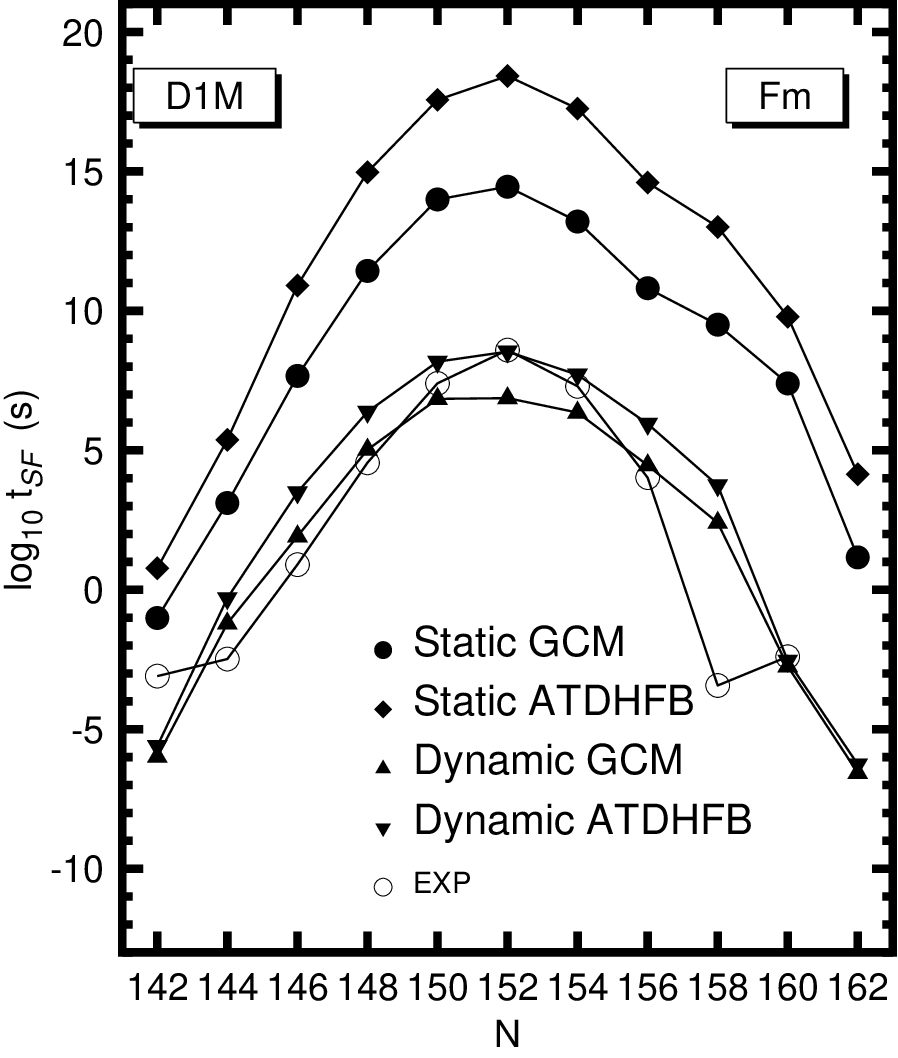} 
\caption{The spontaneous fission half-lives predicted 
within the dynamic GCM and ATDHFB schemes for the isotopes 
$^{242-262}$Fm are depicted as functions of the neutron number N. 
Calculations have been carried out with E$_{0}$ = 0.5 MeV. Results 
corresponding to the static GCM and ATDHFB schemes \cite{Rayner-Pu} are 
also included in the plot. The experimental $t_{SF}$ values are taken 
from Ref.~\cite{Holden-tsf-exp}. } \label{tsf-syst-Fm} 
\end{figure}

The behavior of ${\cal{S}}(Q_{20})$ Eq.(\ref{Integrand-Action}) as a
function of $\langle \Delta \hat{N}^2 \rangle$ for given values of
$Q_{20}$ = 14, 26, 44 and 70 b is plotted 
in panels (d) and (e) of Fig.~\ref{fig-conjunto-E-BATD-BGCM}. Both the 
GCM and ATDHFB actions are given. As can be seen, 
the integrand displays a minimum at a value $\langle \Delta \hat{N}^2 
\rangle$ substantially larger than the one corresponding to the self-consistent 
minimal energy solution, i.e., $\langle \Delta \hat{N}^2 
\rangle_{self}$. For example, in the  
$Q_{20}$ = 44 b case, $\langle \Delta \hat{N}^2 
\rangle_{self}$ = 14 and ${\cal{S}}_{GCM}(Q_{20} = 44 b)$ = 4.8757 
$\times$ 10$^{-3}$  $\hbar$. As a function of $\langle \Delta \hat{N}^2 \rangle$, the minimum value of 
${\cal{S}}_{GCM}(Q_{20} = 44 b)$ turns out to be 3.7799 $\times$ 
10$^{-3}$ $\hbar$ and corresponds to $\langle \Delta \hat{N}^2 \rangle$ 
= 26. In the ATDHFB case, ${\cal{S}}_{ATDHFB}(Q_{20} = 44 b)$ = 5.9091 
$\times$ 10$^{-3}$ $\hbar$ for $\langle \Delta \hat{N}^2 
\rangle_{self}$ = 14 while, the minimum value of 
${\cal{S}}_{ATDHFB}(Q_{20} = 44b)$ turns out to be 4.3781 10$^{-3}$ 
$\hbar$ for $\langle \Delta \hat{N}^2 \rangle$ = 27. This kind of 
quenching, within both the GCM and ATDHFB schemes, reduces considerably 
the action $S$ Eq.(\ref{Action}) that appears  in the exponential of the 
spontaneous fission half-life Eq.(\ref{TSF-WKB}). The impact on the 
predicted $t_{SF}$ values is therefore of exponential character. For example, in the case of 
$^{252}$No, the least action approach leads to the GCM and ATDHFB 
values $\log_{10} t_{SF}$ = 3.1819 ($t_{SF}$ in s) and $\log_{10} t_{SF}$ = 3.9838, 
respectively. They should be compared with the values $\log_{10}
t_{SF}$ = 9.3878  and $\log_{10} t_{SF}$ =11.6662  obtained within 
the static GCM and ATDHFB schemes \cite{Rayner-No}. The strong 
reduction observed in the  dynamic GCM and ATDHFB  spontaneous fission 
half-lives, brings them in closer agreement with the experimental value 
\cite{Holden-tsf-exp}.

One important characteristic of the dynamic calculation is its dependence with
the parameter $E_0$ entering the definition of the action Eq. (\ref{Integrand-Action}).
Its role is different from the one of the static case, where $E_0$
only serves to set up the integration interval in the action.
The parameter $E_{0}$  can be estimated 
using the curvature of the energy around the ground state minimum of 
the  fission path and the values of the collective inertias 
\cite{Baran-SF-2012,Rayner-Ra}. However, we have followed the usual 
recipe of considering it as a free parameter. In particular, the results 
already discussed have been obtained with  $E_{0}$ = 0.5 MeV, a value 
already employed in previous studies 
\cite{Robledo-Giulliani,WERP02,Rayner-U,Rayner-Pu,Rayner-Ra,Rayner-No,Rayner-U_Pu-odd}. 
Nevertheless, calculations have also been carried out with E$_{0}$ = 1.0 MeV 
\cite{Robledo-Giulliani,Rayner-U,Rayner-Pu,Rayner-Ra,Rayner-No,Rayner-U_Pu-odd,Min_Action_RMF}
to test the sensibility of $t_{SF}$ to this parameter. 
The least action spontaneous fission half-lives 
for E$_{0}$ = 1.0 MeV are 
$\log_{10} t_{SF}$ = 1.4302 s (GCM) and $\log_{10} t_{SF}$ = 2.1485 s (ATDHFB).
Therefore, as in the static approach 
\cite{Robledo-Giulliani,Rayner-U,Rayner-Pu,Rayner-Ra,Rayner-No,Rayner-U_Pu-odd}, 
increasing E$_{0}$ (from 0.5 to 1.0 MeV) also provides a reduction 
in the dynamic $t_{SF}$ values. This might lead to a better comparison 
with the experiment \cite{Holden-tsf-exp}. However, as all the trends 
with neutron number obtained in this study remain qualitatively the same regardless of the 
E$_{0}$ value used, in what follows we will restrict our discussions 
(see, Sec.~\ref{systematics-paths}) to those results obtained with E$_{0}$ = 
0.5 MeV. 

 
\subsection{Systematic of the dynamic fission paths 
and spontaneous fission half-lives in $^{242-262}$Fm  and $^{250-260}$No}
\label{systematics-paths}

In Figs.~\ref{paths-242-246-250-254-258-260Fm} and 
\ref{paths-248-252-256-260No}, we have plotted the HFB plus the 
zero-point rotational energies corresponding to the dynamic ATDHFB and 
GCM fission paths, as functions of the quadrupole moment $Q_{20}$, for 
the nuclei $^{242,246,250,254,258,262}$Fm and $^{250,252,256,260}$No. A 
similar pattern is exhibited by other Fm and No nuclei and due to this, 
they are not shown in the figures. Starting from the nuclei $^{246}$Fm 
and $^{252}$No, the curves have been shifted by 50 MeV in order to 
accommodate them in a single plot. The energies corresponding to the 
static path are also included in the plot. We have followed the same 
methodology described in Sec.~\ref{strategy-252No} for $^{252}$No to 
compute the dynamic paths shown in the figures.

Previous studies within the static  framework have pointed out the role 
of triaxiality for configurations around the top of the inner barrier 
(see, for example, \cite{Abusara-2010,Delaroche-2006,Rayner-U}). 
Typically, triaxiality reduces the height of the inner barrier by a few 
MeV. However, the lowering of the inner barrier comes together with an 
increase of the  collective inertia \cite{Bender-1998,Baran-1981} that 
tends to compensate in the final value of the action. Therefore, the 
impact of triaxiality in the  of $t_{SF}$ value is expected to be very limited and it 
has not been considered in this study. Previous 
studies \cite{Baran-1981} analyzing the dynamic path to fission have 
corroborated the insignificant role played by triaxiality to determine 
lifetimes. Moreover, it has also been shown recently, that pairing 
fluctuations can  restore axial symmetry along the fission path 
\cite{Sad14,Min_Action_Skyrme_2}.

The absolute minima of the static and dynamic paths, shown in 
Fig.~\ref{paths-242-246-250-254-258-260Fm}, for Fm isotopes correspond 
to $Q_{20}$ = 12-16 b while  the top of the inner barriers corresponds 
to  $Q_{20}$ = 28-36 b. Fission isomers, located around $Q_{20}$ = 
50-54 b, are apparent from the  paths  of $^{242,246,250,254}$Fm. They 
are less well defined for isotopes with larger neutron numbers. Similar 
features are also observed for the static and dynamic  paths of the 
isotopes $^{250,252,256,260}$No, shown in 
Fig.~\ref{paths-248-252-256-260No}. The sectors 
of the 
fission paths shown in Figs.~\ref{paths-242-246-250-254-258-260Fm} and 
\ref{paths-248-252-256-260No} where octupole correlations 
play a key role correspond to  quadrupole moments $Q_{20} 
\ge$ 60 b. On the other hand, the most pronounced differences between 
the static and dynamic paths are found for the excitation energies of 
the spherical configurations as well as for the inner barrier heights. 
For example, the  inner barrier heights $B_{I}$ corresponding to the 
dynamic GCM and ATDHFB fission paths in $^{242-262}$Fm  and 
$^{250-260}$No are plotted in panels (a) and (b) of 
Fig.~\ref{BI-syst-Fm-No} as functions of the neutron number N. The 
inner barrier heights corresponding to the static paths in those nuclei 
are also included in the plot. In the case of Fm isotopes, the maximum 
of $B_{I}$ is reached at N = 150  for both the static and dynamic 
paths. The same is also true for the dynamic paths in  No isotopes but, 
in this case, the maximum static $B_{I}$ value corresponds to N = 152. 
For both isotopic chains the largest  $B_{I}$  values are the dynamic 
ATDHFB ones. Let us stress that, as already discussed in 
Sec.~\ref{strategy-252No}, larger dynamic barrier heights arise from 
the fact that at the $Q_{20}$-configurations  corresponding to the top 
of the inner barriers the energies display an almost parabolic behavior 
as functions of the particle number fluctuations $\langle \Delta 
\hat{N}^2 \rangle$. This combined with a reduction of the collective 
inertias leads to a minimum of the action for those 
$Q_{20}$-constrained configurations.


\begin{figure} 
\includegraphics[width=0.46\textwidth]{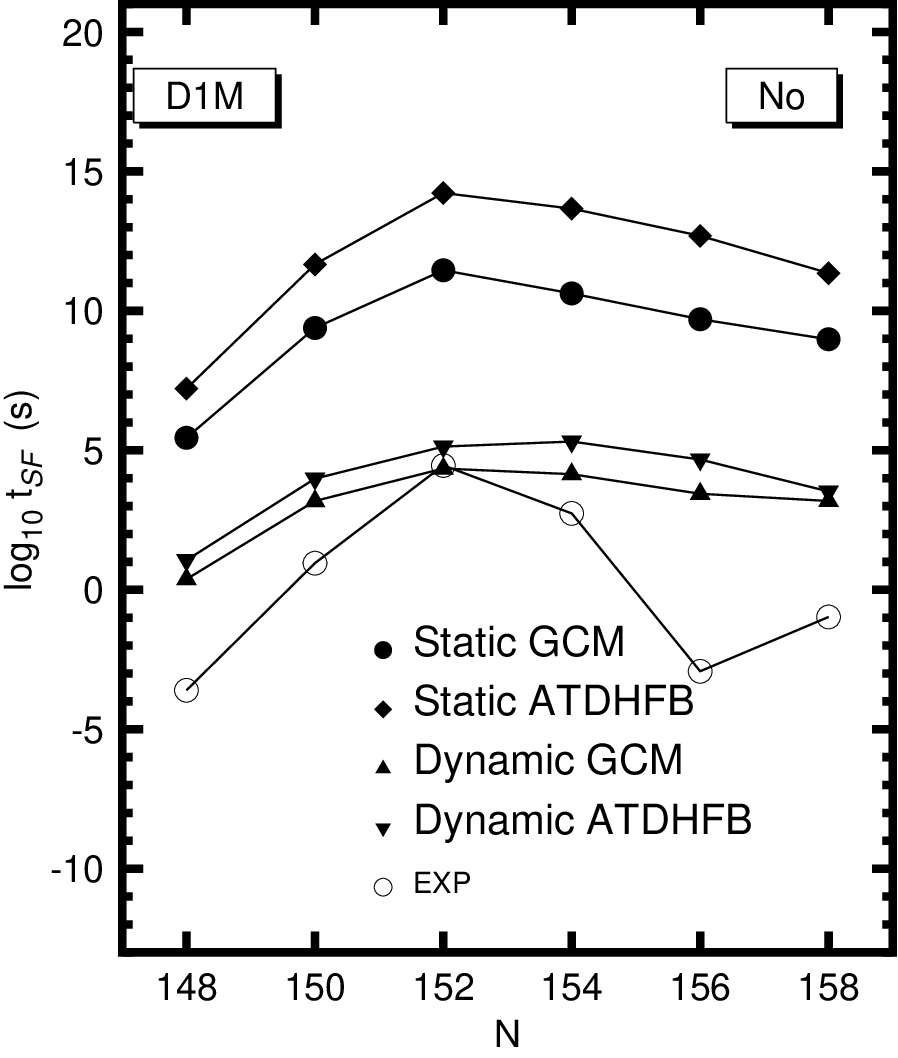} 
\caption{The spontaneous fission half-lives predicted 
within the dynamic GCM and ATDHFB schemes for the isotopes 
$^{250-260}$No are depicted as functions of the neutron number N. 
Calculations have been carried out with E$_{0}$ = 0.5 MeV. Results 
corresponding to the static GCM and ATDHFB schemes \cite{Rayner-No} are 
also included in the plot. The experimental $t_{SF}$ values are taken 
from Ref.~\cite{Holden-tsf-exp}. } \label{tsf-syst-No} 
\end{figure}

In Figs.~\ref{tsf-syst-Fm}    and \ref{tsf-syst-No}, the spontaneous 
fission half-lives predicted within the dynamic GCM and ATDHFB schemes 
are depicted as functions of the neutron number for the nuclei 
$^{242-262}$Fm and $^{250-260}$No. They are compared with the available 
experimental data \cite{Holden-tsf-exp}. Results corresponding to the 
static GCM and ATDHFB schemes  are also 
included in the plot for the sake of comparison. On the one hand, the 
static approximation already accounts qualitatively for the 
experimental bell-shaped dependence of the spontaneous fission 
half-lives as functions of the neutron number \cite{Holden-tsf-exp} in  
Fm isotopes. However, the least action framework provides, via larger 
pairing correlations, a reduction of several orders of magnitude in 
the predicted (dynamic) GCM and/or ATDHFB $t_{SF}$ values.  For 
example, in the case of $^{252}$Fm the static GCM and ATDHFB schemes 
lead to the values $t_{SF}$ = 2.3950 $\times$ 10$^{14}$ s and $t_{SF}$ 
= 2.5686 10$^{18}$ s, respectively \cite{Rayner-Pu}. On the other hand 
the dynamic GCM and ATDHFB values are $t_{SF}$ = 0.7492 $\times$ 
10$^{7}$ s and $t_{SF}$ = 0.3456 $\times$ 10$^{9}$ s, i.e., the least 
action approach provides reductions of 7 and 9 orders of magnitude in 
the predicted spontaneous fission half-lives. As can be seen from 
Fig.~\ref{tsf-syst-Fm}, such a reduction of several orders of magnitude 
occurs for all the considered Fm isotopes and improves dramatically the 
comparison with the experiment. The same holds true for the No isotopes
displayed in Fig. 8, except for the curvature of the experimental values 
around N = 152 that is not yet fully accounted for. 
Another interesting feature of the results shown in the figures is that 
the predictions of the  static GCM and ATDHFB 
schemes  tend to get closer within the dynamic framework. These 
results, and the ones obtained in our previous  study 
\cite{Min_Action_Gogny}, suggest that the least action framework might 
be considered a reasonable starting point to improve the predictive 
power of the HFB approach when applied to the computation of 
spontaneous fission half-lives.

Other calculations of $t_{SF}$ using the Gogny-D1S 
\cite{WERP02} or the Skyrme-SkM* EDFs 
\cite{Baran-SF-2012} in the static approach lead to a good agreement 
with the experimental data. This is not the case for our static results 
with D1M. The reason for the discrepancy with D1S is the 
smaller collective inertia as compared to the D1M one and consequence 
of a larger pairing strength in D1S \cite{Rayner-U}. In the Skyrme-SkM* 
case, the pairing strength is adjusted to the pairing gaps in 
$^{252}$Fm and therefore, it is very likely that it includes in an 
effective way other effects beyond the HFB theory that give  inertia 
values better suited to reproduce the experimental $t_{SF}$.


\section{Conclusions}
\label{Coclusions}

In this paper, we have considered a least action approach to compute 
the spontaneous fission half-lives  $t_{SF}$ for a selected set of 
fermium and nobelium nuclei. To this end, calculations have been 
carried out within the constrained HFB method and the Gogny-D1M EDF. 
The axially symmetric quadrupole moment $Q_{20}$  and the 
particle number fluctuation $\langle \Delta \hat{N}^2 \rangle$ have 
been identified as the relevant degrees of freedom for the minimization 
of the  WKB action. The parabolic behavior of the energy as a function 
of $\langle \Delta \hat{N}^2 \rangle${} together with the corresponding 
decrease of the (GCM and/or ATDHFB) collective masses leads to a 
minimum of the action at a $\langle \Delta \hat{N}^2 \rangle$ value 
larger than the selfconsistent one. As a consequence, the dynamic  GCM 
and/or ATDHFB fission paths exhibit major differences with respect to 
the static ones for both the spherical configuration as well as for 
configurations around the top of the inner barriers for all the nuclei 
studied. As functions of the neutron number, the maxima of the inner 
barrier heights correspond to N $\approx$ 152. Moreover, the larger 
pairing correlations of the dynamic path with the subsequent reduction 
of the action, provide dynamic GCM and ATDHFB spontaneous fission 
half-lives which are several orders of magnitude smaller than the 
corresponding static predictions improving dramatically the comparison 
with the available experimental data. It is 
also found, that the values of $t_{SF}$  
predicted  within the  static GCM and ATDHFB schemes tend to get closer 
within the dynamic framework. These findings are in line with previous 
studies in Ref.~\cite{Min_Action_Gogny}. The pairing enhancement along the 
dynamic path could be simulated in an effective way by increasing the 
pairing strength parameter of the interaction \cite{Rayner-U}. 

A long list of task remains to be undertaken. For example, in this work 
we have resorted to a single constrain on the operator $\Delta 
\hat{N}^2$ associated with the (total) particle number fluctuations. A 
more realistic approach could   be to consider separate constrains  
$\Delta \hat{Z}^2$ and $\Delta \hat{N}^2$ on the proton 
and neutron number fluctuations, respectively. We have considered both 
the GCM and ATDHFB collective inertias within the perturbative cranking 
approximation. The use of a nonperturbative approach for the 
computation of the collective inertias as well as the coupling with 
degrees of freedom other than the quadrupole one also remain to be 
explored within the employed least action framework. Work along these 
lines is in progress and will be reported elsewhere.

\begin{acknowledgments}
The work of LMR was partly supported by Spanish MINECO grant Nos
FPA2015-65929 and FIS2015-63770.

\end{acknowledgments}



\begin{thebibliography}{00}

\bibitem{Bjor} S. Bj\"ornholm and J.E. Lynn, Rev. Mod. Phys. {\bf{52}}, 725 (1980).

\bibitem{Specht} H.J. Specht, Rev. Mod. Phys. {\bf{46}}, 773 (1974).

\bibitem{Krappe} H.J. Krappe and K. Pomorski, {\it{Theory of Nuclear Fission}}, Lectures  Notes
in Physics, {\bf{838}} (2012).

\bibitem{Schunck2016} N. Schunck and L. M. Robledo, Rep. Prog. Phys. {\bf{79}}, 116301 (2016).

\bibitem{Baran2015} A. Baran, M. Kowal, P. -G. Reinhard, 
L. M. Robledo, A. Staszczak and M. Warda, Nucl. Phys. A {\bf 994}, 442 (2015).

\bibitem{naza2018} W. Nazarewicz, Nature Physics, \textbf{14}, 537 (2018).

\bibitem{Holden-tsf-exp} N. E. Holden and D. C. Hoffman, Pure Appl. Chem. {\bf{72}}, 1525 (2000).


\bibitem{rs} P. Ring and P. Schuck, {\em The Nuclear Many-Body Problem} (Springer, 
Berlin, 1980).

\bibitem{gogny-d1s} J. F. Berger, M. Girod, and D. Gogny, Nucl. Phys. A {\bf{428}}, 23c (1984).

\bibitem{Delaroche-2006} J.-P. Delaroche, M. Girod, H. Goutte and J. Libert, Nucl. Phys. A {\bf{771}}, 103 (2006).

\bibitem{Robledo-Martin} V. Martin and L.M. Robledo, Int. J. Mod. Phys. E {\bf{18}}, 788 (2009).

\bibitem{Dubray} N. Dubray, H. Goutte and J.-P. Delaroche, Phys. Rev. C {\bf{77}}, 014310 (2008).

\bibitem{PEREZ-ROBLEDO} S. P\'erez-Mart\'in and L.M. Robledo, Int. J. Mod. Phys. E {\bf{18}}, 861 (2009).

\bibitem{Younes2009} W. Younes and D. Gogny, Phys. Rev. C {\bf{80}}, 054313 (2009).

\bibitem{WERP02} M. Warda, J. L. Egido, L.M. Robledo 
and K. Pomorski, Phys. Rev. C {\bf 66}, 014310 (2002).

\bibitem{Egido-other1} J.L. Egido and L.M. Robledo, Phys. Rev. Lett. {\bf{85}}, 1198
(2000).

\bibitem{Warda-Egido-2012} M. Warda and J.L. Egido, Phys. Rev. C {\bf{86}}, 014322
(2012).

\bibitem{UNEDF1} N. Nikolov, N. Schunck, W. Nazarewicz, M. Bender
 and J. Pei, Phys. Rev. C {\bf{83}}, 034305 (2011).

\bibitem{Mcdonell-2} J.D. McDonnell, W. Nazarewicz and J.A. Sheikh, Phys. Rev. C {\bf{87}}, 054327 (2013).


\bibitem{Erler2012} J. Erler, K. Langanke, H.P. Loens, G. Mart\'inez-Pinedo 
and P.-G. Reinhard, Phys. Rev. C {\bf{85}}, 025802 (2012).

\bibitem{Baran-SF-2012} A. Staszczak, A. Baran, W. Nazarewicz, Phys. Rev. C {\bf{87}}, 024320 (2013) 

\bibitem{Baran-1981} A. Baran, K. Pomorski, A. Lukasiak and A. Sobiczewski, Nucl. Phys. A {\bf{361}}, 83 (1981).

\bibitem{BCPM} M. Baldo, L.M. Robledo, P. Schuck and X. Vi\~nas, Phys. Rev. C {\bf{87}}, 064305 (2013).

\bibitem{Robledo-Giulliani} S.A. Giuliani and L.M Robledo, Phys. Rev. C {\bf{88}}, 054325 (2013).

\bibitem{Giuliani2018} Samuel A. Giuliani, Gabriel Mart\'\i nez-Pinedo, and Luis M. Robledo
Phys. Rev. C {\bf 97}, 034323 (2018)


\bibitem{Abusara-2010} H. Abusara, A.V. Afanasjev and P. Ring, Phys. Rev. C {\bf{82}}, 044303 (2010).

\bibitem{Abu-2012-bheights} H. Abusara, A.V. Afanasjev and  P. Ring, Phys. Rev. C {\bf 85}, 024314 (2012).

\bibitem{RMF-LU-2012} B.-N. Lu, E.-G. Zhao and S.-G. Zhou, Phys. Rev. C {\bf{85}}, 011301 (2012).

\bibitem{Kara-RMF} S. Karatzikos, A. V. Afanasjev, G. A. Lalazissis and  P. Ring, Phys. Lett. B {\bf{689}}, 72 (2010).

\bibitem{Bender-1998} 
M. Bender, K. Rutz, P.-G. Reinhard, J.A. Maruhn and W. Greiner, 
Phys. Rev. C {\bf 58}, 2126 (1998). 

\bibitem{Buervenich2004} T. B\"urvenich, M. Bender, J. A. Maruhn, and P.-G. Reinhard,
Phys. Rev. C {\bf 69}, 014307 (2004)

\bibitem{Rutz1997} K. Rutz, M. Bender, T. B\"urvenich, T. Schilling, P.-G. Reinhard, J. A. Maruhn, and W. Greiner
Phys. Rev. C {\bf 56}, 238 (1997)


\bibitem{Rayner-U} R. Rodr\'iguez-Guzm\'an and L. M. Robledo, Phys. Rev. C {\bf{89}}, 054310 (2014).

\bibitem{Rayner-Pu} R. Rodr\'iguez-Guzm\'an and L. M. Robledo, Eur. Phys. J. A {\bf{50}}, 142 (2014).

\bibitem{Rayner-Ra} R. Rodr\'iguez-Guzm\'an and L. M. Robledo, Eur. Phys. J. A {\bf{52}}, 12 (2016).


\bibitem{gogny-d1m} S. Goriely, S. Hilaire, M. Girod 
and S. P\'eru, Phys. Rev. Lett. {\bf 102}, 242501 (2009).

\bibitem{gogny-d1n} F. Chappert, M. Girod, and S. Hilaire, Phys. Lett. B {\bf{668}}, 420 (2008).

\bibitem{Baran-mass-2011} A. Baran, J.A. Sheikh, J. Dobaczewski, W. Nazarewicz 
and A. Staszczak, Phys. Rev. C {\bf{84}}, 054321 (2011).

\bibitem{Baran-tsf-other} A. Baran, Phys. Lett. B {\bf{76}}, 8 (1978).

\bibitem{Rayner-No} R. Rodr\'iguez-Guzm\'an and L. M. Robledo, Eur. Phys. J. A {\bf{52}}, 348 (2016).

\bibitem{Rayner-U_Pu-odd} R. Rodr\'iguez-Guzm\'an and L. M. Robledo, Eur. Phys. J. A {\bf{53}}, 245 (2017).

\bibitem{Sara-Robledo_EFA} S. P\'erez-Mart\'in and L.M. Robledo, Phys. Rev. C {\bf{78}}, 014304 (2008).


\bibitem{Min_Action_Skyrme_1} J. Sadhukhan, K. Mazurek, A. Baran, J. Dobaczewski, W. Nazarewicz and
J. A. Sheikh, Phys. Rev. C {\bf{88}}, 064314 (2013).

\bibitem{Min_Action_Gogny} S.A. Giuliani, L. M. Robledo and R. Rodr\'iguez-Guzm\'an, Phys. Rev. C {\bf{90}}, 054311 (2014).

\bibitem{Moreto-pairing-fission} L. G. Moretto and R. P. Babinet, Phys. Lett. B {\bf{49}}, 147 (1974).

\bibitem{proportional-1} M. Brack, J. Damgaard, A.S. Jensen, H.C. Pauli, V.M Strutinsky
and C.Y. Wong, Rev. Mod. Phys. {\bf{44}}, 320 (1972).

\bibitem{proportional-2} G.F. Bertsch and H. Flocard, Phys. Rev. C {\bf{43}}, 2200 (1991).

\bibitem{Urin} M. Urin and D. Zaretsky, Nucl. Phys. {\bf{75}}, 101 (1976).


\bibitem{Lublin-1} A. Staszczak, A. Baran, K. Pomorski and K. B\"oning, Phys. Lett. B {\bf{161}}. 

\bibitem{Lublin-2} K. Pomorski. Int. J. of Mod. Phys. E {\bf{16}}, 237 (2007). 

\bibitem{Sta-P2} A. Staszczak, S. Pilat and K. Pomorski, Nucl. Phys. A {\bf{504}}, 589 (1989).

\bibitem{Sad14} J. Sadhukhan, J. Dobaczewski, W. Nazarewicz, J. A. Sheikh, and A. Baran,
                Phys. Rev. C \textbf{90}, 061304(R) (2014). 

\bibitem{Min_Action_Skyrme_2} J. Sadhukhan, W. Nazarewicz and
N. Schunck, Phys. Rev. C {\bf{93}}, 011304 (2016).


\bibitem{Min_Action_RMF} J. Zhao, B. -N. Lu, T. Niksic, D. Vretenar 
and S. -G. Zhou, Phys. Rev. C {\bf{93}}, 044315 (2016).



\bibitem{Ber91} J.F.Berger, M.Girod, D.Gogny, Comp. Phys. Comm. \textbf{63}, 365 (1991).

\bibitem{Gou05} H. Goutte, J. F. Berger, P. Casoli, and D. Gogny, Phys. Rev. C \textbf{71}, 024316 (2005).

\bibitem{Reg16} D. Regnier, N. Dubray, N. Schunck, and M. Verriere, Phys. Rev. C \textbf{93}, 054611 (2016).

\bibitem{Tao17} H. Tao, J. Zhao, Z. P. Li, T. Niksic, and D. Vretenar, Phys. Rev. C \textbf{96}, 024319 (2017).

\bibitem{Zdeb17} A. Zdeb, A. Dobrowolski, and M. Warda, Phys. Rev. C \textbf{95}, 054608 (2017).

\bibitem{Tan17} Y. Tanimura, D. Lacroix, and S. Ayik, Phys. Rev. Lett. \textbf{118}, 152501 (2017)

\bibitem{God15} P. Goddard, P. Stevenson, and A. Rios, Phys. Rev. C \textbf{92}, 054610 (2015).

\bibitem{Bul16} A. Bulgac, P. Magierski, K. J. Roche, and I. Stetcu, Phys. Rev. Lett. \textbf{116}, 122504 (2016)

\bibitem{Sim18} C. Simenel, A.S. Umar, arXiv:1807.01859



\bibitem{PRCQ2Q3-2012} 
R. Rodr\'iguez-Guzm\'an, L.M. Robledo and P. Sarriguren, 
Phys. Rev. C {\bf 86}, 034336 (2012).

\bibitem{PTpaper-Rayner} R. Rodr\'iguez-Guzm\'an, P. Sarriguren,  L.M. Robledo,
and J. E. Garc\'ia-Ramos, Phys. Rev. C {\bf 81}, 024310 (2010).

\bibitem{Rayner-Sara} R. Rodr\'iguez-Guzm\'an, P. Sarriguren, L.M. Robledo, and
S. Perez-Martin, Phys. Lett. B {\bf{691}}, 202 (2010).

\bibitem{Rayner-Robledo-JPG-2009} L.M. Robledo, R. Rodr\'iguez-Guzm\'an, and P. Sarriguren, J.
Phys. G: Nucl. Part. Phys. {\bf{36}}, 115104 (2009).

\bibitem{Robledo-Rayner-JPG-2012} L.M. Robledo and R. Rodr\'iguez-Guzm\'an, J.
Phys. G: Nucl. Part. Phys. {\bf{39}}, 105103 (2012).


\bibitem{Rayner-PRC-2010} R. Rodr\'iguez-Guzm\'an, P. Sarriguren, and L.M. Robledo, Phys.
Rev. C {\bf{82}}, 044318 (2010).


\bibitem{Rayner-PRC-2011} R. Rodr\'iguez-Guzm\'an, P. Sarriguren, and L.M. Robledo, Phys.
Rev. C {\bf{83}}, 044307 (2011).



\bibitem{Robledo-Bertsch2OGM} L.M. Robledo and G. F. Bertsch, Phys. Rev. C {\bf{84}}, 014312 (2011).




\bibitem{CoulombSlater} 
C. Titin-Schnaider and Ph. Quentin, 
Phys. Lett. B {\bf{49}}, 213 (1974).

\bibitem{crankingAPPROX} 
M. Girod and B. Grammaticos, 
Nucl. Phys. A {\bf{330}}, 40 (1979).


\bibitem{Giannoni} 
M.J. Giannoni  and P. Quentin, 
Phys. Rev. C {\bf{21}}, 2060 (1980); 
Phys. Rev. C {\bf{21}}, 2076 (1980).

\bibitem{Libert-1999} 
J. Libert, M. Girod and  J.P. Delaroche,  
Phys. Rev. C {\bf{60}}, 054301 (1999).


\bibitem{ER-Lectures} 
J.L. Egido and L.M.Robledo, 
Lectures Notes in Physics {\bf{641}}, 269 (2004).

\bibitem{RRG23S} 
R. Rodr\'iguez-Guzm\'an, J.L. Egido and  L.M. Robledo, 
Phys. Lett. B {\bf{474}}, 15 (2000); 
Phys. Rev. C {\bf 62}, 054308 (2000).

\bibitem{NPA-2002} 
R. Rodr\'iguez-Guzm\'an, J.L. Egido, and L.M. Robledo,
Nucl. Phys. A {\bf{709}}, 201 (2002).





\bibitem{Bertsch2015} G.F. Bertsch, W. Loveland, W. Nazarewicz,  and P. Talou, 
J. Phys. G Nucl. Part. Phys. {\bf 42}, 077001 (2015)



\end{thebibliography}
\end{document}